# Optical Cloaking with Non-Magnetic Metamaterials

Wenshan Cai, Uday K. Chettiar, Alexander V. Kildishev and Vladimir M. Shalaev

*School of Electrical and Computer Engineering and Birck Nanotechnology Center, Purdue University, West Lafayette, Indiana 47907, USA*

**Abstract:** Artificially structured metamaterials have enabled unprecedented flexibility in manipulating electromagnetic waves and producing new functionalities, including the cloak of invisibility based on coordinate transformation. Here we present the design of a *non-magnetic* cloak operating at optical frequencies. The principle and structure of the proposed cylindrical cloak are analyzed, and the general recipe for the implementation of such a device is provided. The cloaking performance is verified using full-wave finite-element simulations.

The recently proposed electromagnetic cloak based on coordinate transformation has opened a new door for the applications of metamaterials[1,2]. Unlike other cloaking approaches[3,4,5] which are typically limited to sub-wavelength objects, the transformation method allows the design of cloaking devices to render a macroscopic object invisible. In addition, the design is not sensitive to the object that is being cloaked. The first experimental demonstration of such a cloak at microwave frequencies was recently reported[6]. We note, however, that the design used[6] cannot be implemented for an optical cloak, which is certainly of particular interest because optical frequencies are where the word "invisibility" is conventionally defined.

In this paper we present the design of a non-magnetic cloak operating at optical frequencies. The coordinate transformation used in the proposed cloak of cylindrical geometry is similar to that in Ref. 6, by which a cylindrical region $r<b$ is compressed into a concentric cylindrical shell $a<r<b$ as shown in Fig. 1a. This transformation results in the following requirements for the anisotropic permittivity and permeability in the cloaking shell[6,7]:

$$\varepsilon_r = \mu_r = \frac{r-a}{r}, \quad \varepsilon_\theta = \mu_\theta = \frac{r}{r-a}, \quad \varepsilon_z = \mu_z = \left(\frac{b}{b-a}\right)^2 \frac{r-a}{r}. \quad (1)$$

For TE illumination with incident electrical field polarized along the *z* axis, only $\varepsilon_z$, $\mu_r$ and $\mu_\theta$ in equation (1) enter into Maxwell's equations. Moreover, the dispersion properties and wave trajectory in the cloaking shell remain the same as long as the values of $\varepsilon_i \mu_z$ and $\mu_i \varepsilon_z$ are kept constant, where *i* represents either *r* or $\theta$. All these have been addressed in the recent microwave experiments[6], where the cloaking was achieved by varying the dimensions of a series of split ring resonators (SRRs) to yield a desired gradient of permeability in the radial direction. However, this approach cannot be used for an optical cloak. It is a known fact that there are intrinsic limits to the scaling of SRR size in order to exhibit a magnetic response in the optical range[8,9]. Replacing the SRRs with other optical magnetic structures like paired nano-rods[10] or nano-strips[11] is also a very challenging approach primarily due to fabrication difficulties. The layer-by-layer fashion of optical or e-beam lithography is not compatible with making closed surfaces such as a cloak. Moreover, optical magnetism based on such resonant plasmonic structures is usually associated with a high loss factor, which is detrimental to the performance of cloaking devices.

In contrast to the reported design of a microwave cloak with TE polarization[6], we focus on TM incidence with the magnetic field polarized along the $z$ axis. In this case only $\mu_z$, $\varepsilon_r$ and $\varepsilon_\theta$ must satisfy the requirements in (1), and the dispersion relations inside the cloak remain unaffected as long as the products of $\mu_z\varepsilon_r$ and $\mu_z\varepsilon_\theta$ are kept the same as those determined by the values in (1). It is worth noting that unlike the TE case, under TM illumination only one component of $\mu$ is of interest, which allows us to completely remove the need for any optical magnetism. In (1) we multiply $\varepsilon_r$ and $\varepsilon_\theta$ by the value of $\mu_z$ and obtain the following reduced set of cloak parameters:

$$\mu_z = 1, \quad \varepsilon_\theta = \left(\frac{b}{b-a}\right)^2, \quad \varepsilon_r = \left(\frac{b}{b-a}\right)^2 \left(\frac{r-a}{r}\right)^2. \tag{2}$$

Compared to the cloak with ideal properties as shown in (1), the reduced parameters in equation (2) provide the same wave trajectory. The only adverse effect of using the reduced set is that the impedance at the outer boundary is not perfectly matched and hence some reflection will exist.

The non-magnetic nature of the system as indicated in (2) removes the most challenging issue of the design. The azimuthal permittivity $\varepsilon_\theta$ is a constant with a magnitude larger than 1, which can be easily achieved in conventional dielectrics. The key to the implementation is to construct the cylindrical shell with the desired radial distribution of $\varepsilon_r$ varying from 0 at the inner boundary of the cloak ($r = a$) to 1 at the outer surface ($r = b$).

Artificial dielectrics with positive permittivity less than unity were first studied more than half a century ago[12] and are still of interest to metamaterial researchers[13]. In our design, the required distribution of $\varepsilon_r$ is realized by using metal wires of subwavelength size in the radial direction embedded in a dielectric material, as shown in Fig. 1b. The aspect ratio of the metal wires, defined by the ratio of the length to the radius of the wire, is denoted by $\alpha$. The whole structure of the cloaking system resembles a round hair brush, except that in a real device the "bristles" of such a brush can consist of disconnected smaller pieces. The spatial positions of the rods don't have to be periodic and can be random.

The shape-dependent electromagnetic response of a subwavelength particle can be characterized by the Lorentz depolarization factor $q$. For an ellipsoid of semiaxes $a_i$, $a_j$ and $a_k$ with electric field polarized along $a_i$, the depolarization factor is expressed by[14]

$$q_i = \int_0^\infty \frac{a_i a_j a_k ds}{2(s+a_i^2)^{3/2}(s+a_j^2)^{1/2}(s+a_k^2)^{1/2}}. \tag{3}$$

The screening factor $\kappa$ of a particle is related to $q$ by $\kappa = (1-q)/q$. Note that a long wire with large aspect ratio $\alpha$ results in a small depolarization factor and a large screening factor, which indicate strong interactions between the field and the wires. For a composite cloak with metal wires as inclusions in a dielectric, the electromagnetic properties are well-described by "shape-dependent" effective-medium theory (EMT). The effective permittivity $\varepsilon_{eff}$ for a composite material comprising metal particles with permittivity $\varepsilon_m$, a volume filling factor $f$ and screening factor $\kappa$, along with a dielectric component with permittivity $\varepsilon_d$ and a filling factor $1-f$, is given by[15]

$$f \frac{\varepsilon_m - \varepsilon_{eff}}{\varepsilon_m + \kappa \varepsilon_{eff}} + (1-f) \frac{\varepsilon_d - \varepsilon_{eff}}{\varepsilon_d + \kappa \varepsilon_{eff}} = 0. \quad (4)$$

For spherical particles with $q = 1/3$ and $\kappa = 2$, the equation above reduces to the common EMT expression[16]. Equation 4 is a quadratic equation with the following solutions (see, for example Ref. 17):

$$\varepsilon_{eff} = \frac{1}{2\kappa} \left\{ \bar{\varepsilon} \pm \sqrt{\bar{\varepsilon}^2 + 4\kappa \varepsilon_m \varepsilon_d} \right\}, \quad (5)$$

where $\bar{\varepsilon} = [(\kappa+1)f - 1]\varepsilon_m + [\kappa - (\kappa+1)f]\varepsilon_d$. The sign in equation (5) should be chosen such that $\varepsilon''_{eff} > 0$.

The benefit of using metal wires in a composite cloak is that the radial permittivity $\varepsilon_r$ determined by (5) may exhibit a positive value less than 1 with minimal imaginary part. For the structure in Fig. 1b, it is easy to see that the volume filling fraction is inversely proportional to $r$. To be more specific, the filling fraction in (5) for calculating $\varepsilon_r$ is $f(r) = P_a \cdot (a/r)$, with $P_a$ being the surface coverage ratio of metal at the inner surface of the cloak ($r = a$). The filling fractions $f$ at the inner and outer surface of the cloak are $P_a$ and $P_a \cdot (a/b)$ respectively, and the overall metal filling fraction in the whole cloak layer is $P_a \cdot 2a/(a+b)$. The azimuthal permittivity $\varepsilon_\theta$ inside the cloak is essentially the same as that of the dielectric because the response of the wires to the angular electrical field $E_\theta$ oriented normal to the wires is small and at low metal filling factors it can be neglected.

The reduced set of cloak parameters in (2) requires a smooth variation of the radial permittivity from 0 to 1 as $r$ changes from $a$ to $b$. That is, at the operational condition,

$$\begin{cases} \varepsilon_{eff,r}(P_a) = 0 \\ \varepsilon_{eff,r}(P_a \cdot a/b) = 1 \end{cases}. \quad (6)$$

For optimal performance, $\varepsilon_{eff,r}$ should exactly follow the function described in (2) such that $\varepsilon_{eff,r}(P_a \cdot a/r) = [b/(b-a)]^2 [(r-a)/r]^2$. In a practical design, $\varepsilon_{eff,r}$ is allowed to have some discrepancy from the optimal value inside the cloak. The most important points are at the inner and outer surfaces of the cloak, where (6) should be satisfied exactly. This ensures perfect index matching at $r = b$ and the minimum leakage energy at $r = a$.

To determine all of the parameters of the design shown in Fig. 1b, we define two filling fraction functions $f_0(\lambda, \alpha)$ and $f_1(\lambda, \alpha)$ such that for given constituent composite materials and a wire aspect ratio of $\alpha$, the effective radial permittivity is

$$\begin{cases} \varepsilon_{eff,r}(\lambda, f_0(\lambda, \alpha)) = 0 \\ \varepsilon_{eff,r}(\lambda, f_1(\lambda, \alpha)) = 1 \end{cases}. \quad (7)$$

Combining equations (6) and (7), at the operational wavelength $\lambda$ we obtain

$$\begin{cases} f_0(\lambda, \alpha) = P_a \\ f_1(\lambda, \alpha) = P_a \cdot a/b \end{cases}. \quad (8)$$

Let $R_{ab} = a/b$ denote the shape factor of the cylindrical cloak, that is, the ratio between the inner and outer radii. From the above relations we can express $R_{ab}$ as

$$R_{ab} = f_1(\lambda, \alpha) / f_0(\lambda, \alpha). \quad (9)$$

Using equation (9) with the expression for $\varepsilon_\theta$ in (2), we obtain the operating condition of the cloak:

$$\varepsilon_\theta(\lambda) = \left( \frac{f_0(\lambda,\alpha)}{f_0(\lambda,\alpha) - f_1(\lambda,\alpha)} \right)^2, \tag{10}$$

where $\varepsilon_\theta(\lambda)$ is the permittivity of the dielectric material surrounding the metal wires in the cloak.

For practical applications, it is important to design a cloaking device operating at a pre-set operational wavelength $\lambda_{op}$. For this purpose the design process is as follows. First we choose materials for the metal wires and the surrounding dielectric. Second, we calculate the values of $f_0$ and $f_1$ as functions of the aspect ratio $\alpha$ at $\lambda_{op}$ using the EMT model in (5). The required aspect ratio for $\lambda_{op}$ is the one that satisfies equation (10). Then, the geometrical factors of the cloak, including $R_{ab}$ and $P_a$, can be determined based on equations (8) and (9). Note that the same design works for all similar cylindrical cloaks with the same shape factor $R_{ab}$. We emphasize again that such cloaking device can be used for large objects.

As a practical example, we have designed an optical cloak operating at the commonly used wavelength of 632.8 nm (He-Ne laser) and consisting of silver and silica. The equations (5), (7) and (10) together yield the desired aspect ratio $\alpha = 10.7$, and the volume filling fractions at the two boundaries are $f_0 = 0.125$ and $f_1 = 0.039$, respectively. Then with (8) and (9) we find the shape factor of the cylindrical cloak to be $R_{ab} = 0.314$, while the surface coverage ratio at the inner boundary is $P_a = 12.5\%$. The effective parameters of $\mu_z$, $\varepsilon_r$ and $\varepsilon_\theta$ from this design together with the exact set of reduced parameters determined by equation (2) are shown in Fig. 2. We can see that $\mu_z$ and $\varepsilon_\theta$ perfectly match the theoretical requirements throughout the cylindrical cloak. The radial permittivity $\varepsilon_r$ fits the values required by equation (2) exactly at the two boundaries of the cloak, and follows the overall tendency very well inside the cloak.

To validate if the required distribution of permittivity could be achieved using prolate spheroidal silver nanowires embedded in a silica tube, we use three-dimensional full-wave simulations with the commercial finite-element solver COMSOL MULTIPHYSICS to determine the effective anisotropic permittivity of a unit cell with sub-wavelength dimensions. We start with a homogenization method[6], where the actual unit cells (cylindrical sectors) with different electromagnetic surroundings at the inner and outer curved boundaries are substituted by cells made of right rectangular prisms, as shown in Fig. 3. The curvature of the actual unit cells is relatively low, thus we assume that converting the cylindrical segment into the rectangular prism introduces minuscule change to the effective permittivity. For a wavelength of 632.8 nm, we fix two dimensions of the unit cell (height, $h_c = 12.5$ nm and length, $l_c = 100$ nm), while changing the width $w_c$ proportional to the radius of each layer. The full-wave finite-element (FE) numerical analysis confirms that the range of desired $\varepsilon_r'$ and $\varepsilon_\theta'$ agrees well with those predicted by EMT. For the equivalent rectangular unit cell encapsulating a spheroidal silver nanowire with the dimensions initially calculated through EMT (diameter $d = 7$ nm and length $l = 75$ nm), the effective permittivity $\varepsilon_r'$ fits relatively well to the desired values, with a discrepancy of around 10%. The required effective material properties $\varepsilon_r'$ and $\varepsilon_\theta'$ can be achieved precisely by additionally adjusting the diameter of the rod to $d = 6$ nm ($d = 5$ nm), and length to $l = 71$ nm ($l = 60$ nm) for the external (internal) cell. As for the loss feature, the FE simulations show that the radial permittivity $\varepsilon_r$ has an imaginary part of about 0.1 throughout the cloak. Although this is a very small value for metal-dielectric metamaterials, it may still weaken the cloaking effect. It is possible to fully compensate the loss by using a gain medium as already proposed for applications of perfect tunneling transmittance[4] and lossless negative-index materials[18,19].

To illustrate the performance of the proposed non-magnetic optical cloak with a design corresponding to Fig. 2 and operating at $\lambda_{op}$ = 632.8 nm, we performed field mapping simulations using a commercial FE package (COMSOL MULTIPHYSICS). The simulation approach is similar to Ref. 7 but with the important difference that our cloaking device is designed for optical wavelengths with TM incident light instead of the TE mode at microwave frequencies. The object hidden inside the cloaked region is an ideal metallic cylinder with radius $r = a$. The simulation domain also consisted of PML layers at all the boundaries to absorb the outgoing waves. The simulated results of magnetic field distribution around the cloaked object together with the power flow lines are illustrated in Fig. 4. We note that the size of the cloak is more than six times the operational wavelength, while the simulated area is more than 20 times the wavelength. Hence both of these sizes are significantly larger than the wavelength used in the calculations. Fig. 4a shows the field distribution around the metal cylinder surrounded by the designed cloak with parameters given by the diamond markers in Fig. 2. With the cloak (Fig. 4a) the wave fronts flow around the cloaked region with remarkably small perturbation, while without the cloak (Fig. 4b) the waves around the object are severely distorted and an evident shadow is cast behind the cylinder. These simulations clearly show the capability of reducing the scattering from the object hidden inside the cloaked region.

We have demonstrated a design of an optical cloak based on coordinate transformation. The non-magnetic nature of our design eases the pain of constructing gradient magnetic metamaterials in three-dimensional space, and therefore paves the way for the realization of cloaking devices at optical frequencies. The proposed design can be generalized to cloaks with other metal structures, such as chains of metal nanoparticles or thin continuous or semi-continuous metal strips. It can be also adopted for other than the optical spectral ranges, including the infrared and the microwave. We note that the achievable invisibility with the proposed cloak is not perfect due to impedance mismatch associated with the reduced material specifications and the inevitable loss in a metal-dielectric structure. Moreover, any shell-type cloak design can work only over a narrow frequency range, because the curved trajectory of light implies a refractive index $n$ of less than 1 in order to satisfy the minimal optical path requirement of Fermat's principle, while any metamaterial with $n < 1$ must be dispersive to fulfill causality. However, we believe that even rudimentary designs and implementations of an optical cloak are of great potential interest and bring us one step closer to the ultimate optical illusion of invisibility.

**Figure 1**

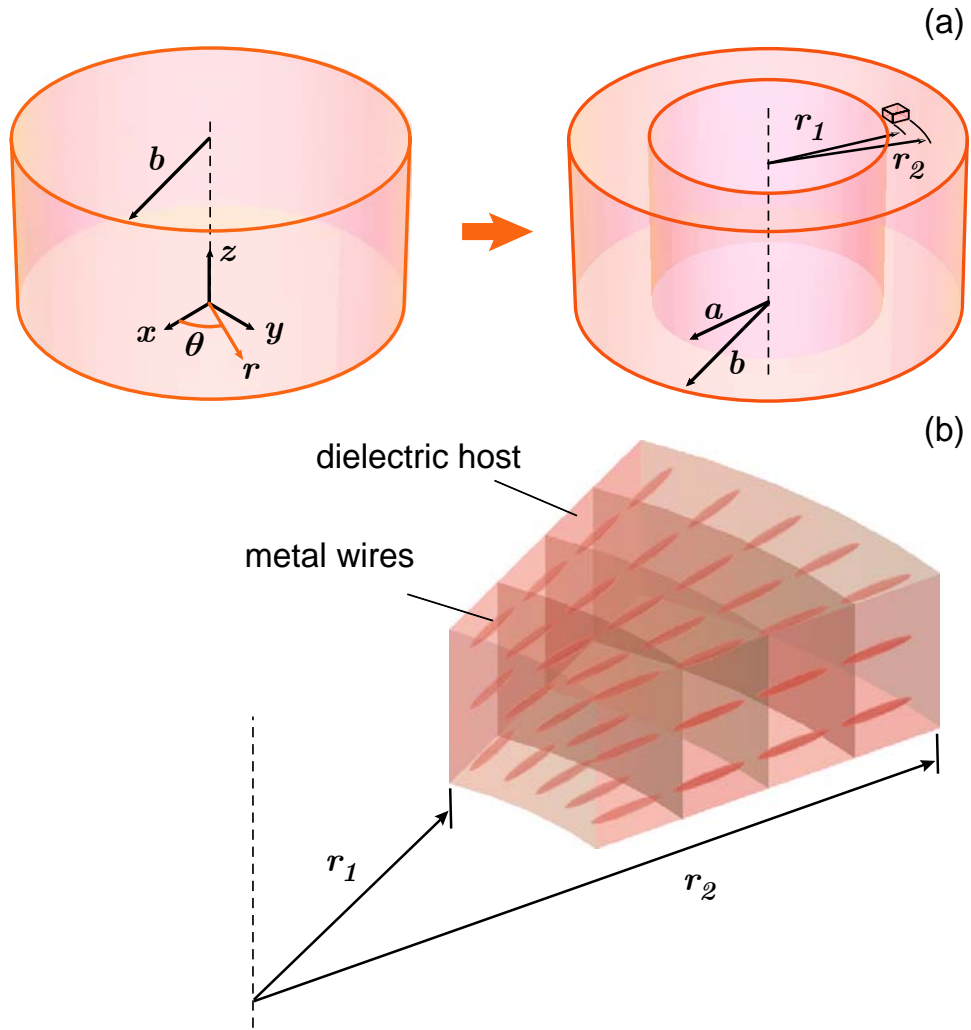

**Figure 1: Coordinate transformation and structure of the designed non-magnetic optical cloak. a,** The coordinate transformation that compresses a cylindrical region $r < b$ into a concentric cylindrical shell $a < r < b$. There is no variation along the $z$ direction. **b,** A small fraction of the cylindrical cloak. The wires are all perpendicular to the cylinder's inner and outer interfaces but their spatial positions don't have to be periodic and can be random. Also, for large cloaks, the wires can be broken into smaller pieces that are smaller in size than the wavelength.

**Figure 2**

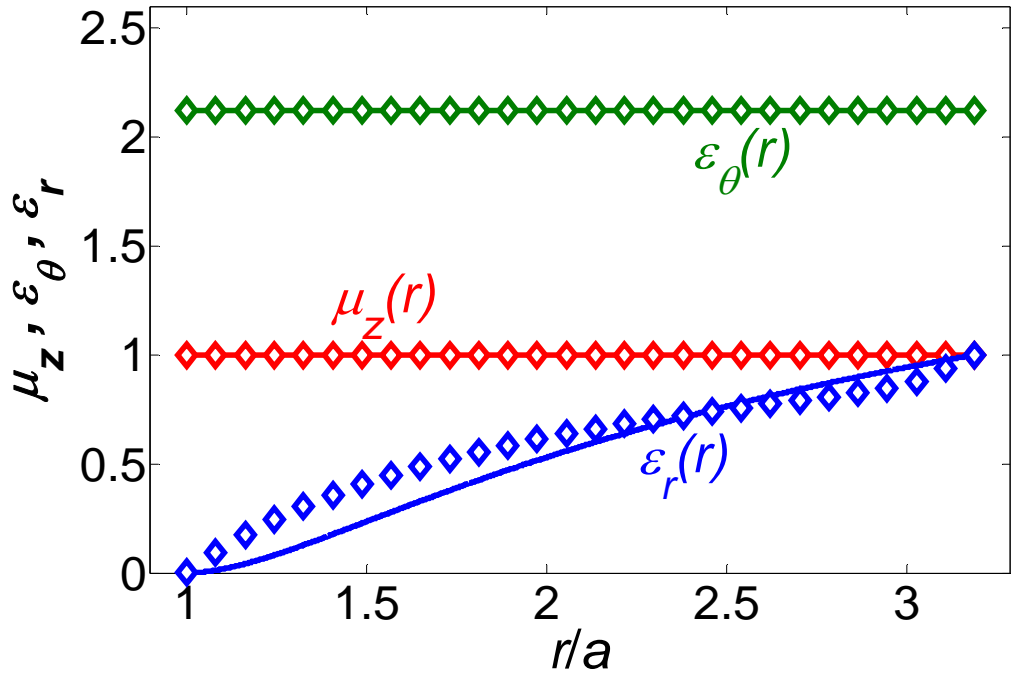

**Figure 2: Material parameters $\varepsilon_r$, $\varepsilon_\theta$ and $\mu_z$ of the proposed cloak operating at $\lambda$ = 632.8 nm.** The solid lines (——) represent the exact set of reduced parameters determined by equation (2). The diamond markers ( ◊ ) show the material properties of the designed metal wire composite cloak with parameters obtained from equations (5) to (10).

**Figure 3**

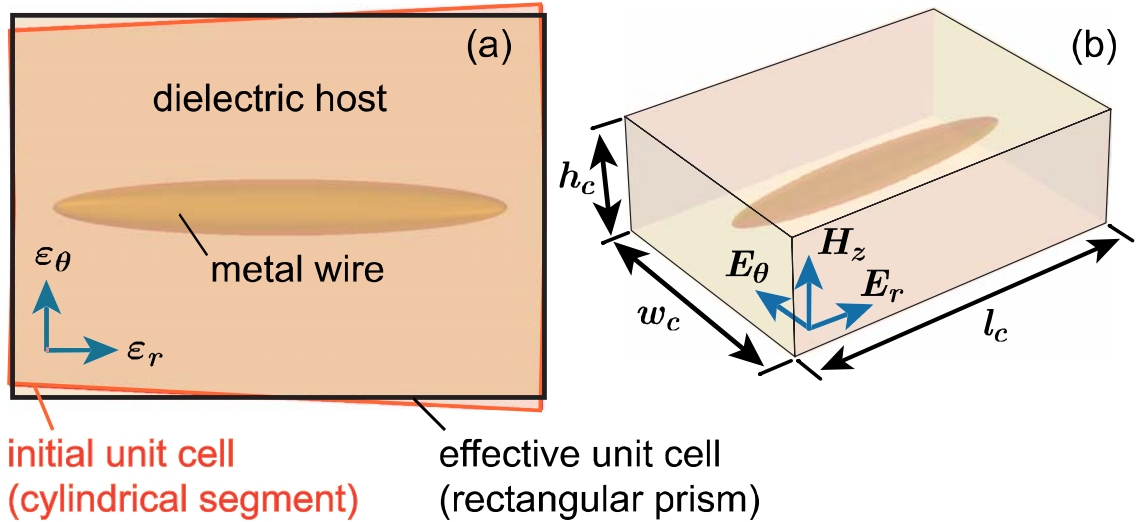

**Figure 3: Unit cell for full-wave finite-element simulations of effective parameters. a,** The actual unit cell (cylindrical sector) encapsulating a spheroidal silver wire is substituted by a cell made of a right rectangular prism. **b,** The geometry of the 3D rectangular unit cell. In simulations $h_c$ and $l_c$ are fixed, while $w_c$ changes in proportion to the radius of each layer.

**Figure 4**

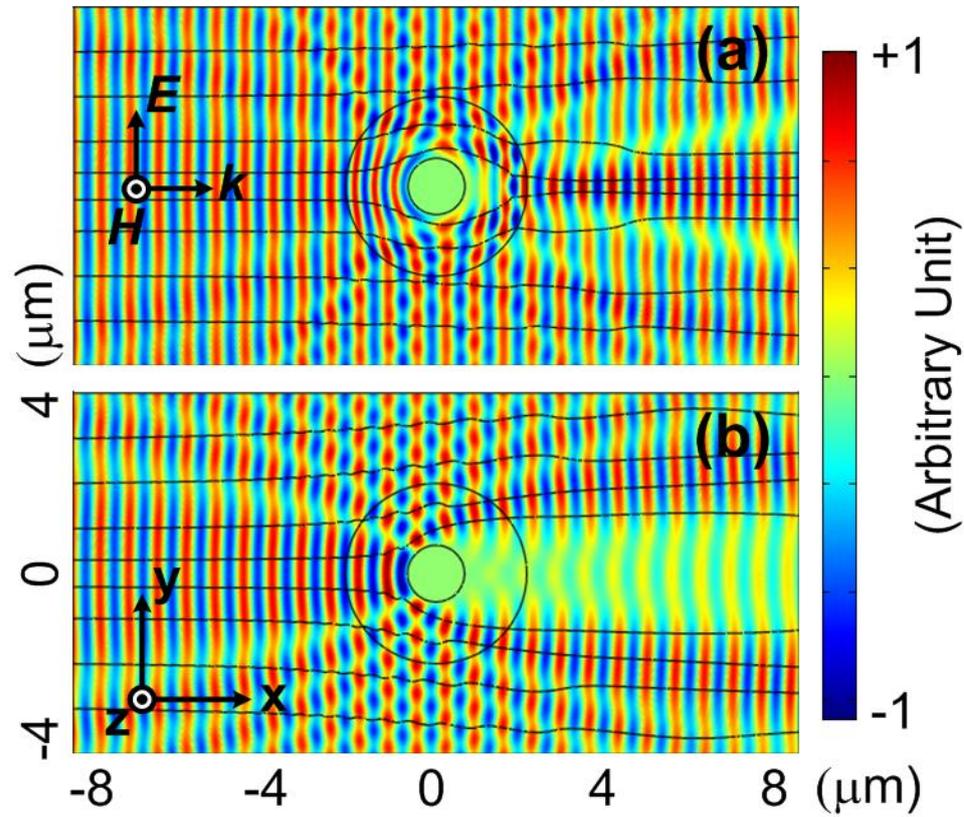

**Figure 4: Finite-element simulations of the magnetic field mapping around the cloaked object with TM illumination at $\lambda$ = 632.8 nm. a,** The object is inside the designed metal wire composite cloak with parameters given by the diamond markers in Fig. 2. **b,** The object is surrounded by vacuum without the cloak. The concentric circles represent the two boundaries of the cloak at $r = a$ and $r = b$, respectively. The hidden object is an ideal metallic cylinder with radius $r = a$.